\documentclass[twocolumn,preprintnumbers,amsmath,amssymb]{revtex4}

\begin{document}
\title{Photon Redshift in a Magnetic field}
\author{ H. P\'erez Rojas and E. Rodriguez Querts}
\affiliation{Instituto de Cibernetica, Matematica y Fisica, Calle E
309, Vedado, Ciudad Habana, Cuba.\\}
\date{\today}

\begin{abstract}
Previous results from the authors concerning the arising a tiny
photon anomalous paramagnetic moment are also interpreted as a
red-shift in analogy to the gravitational known effect. It is due to
the photon interaction with the magnetized virtual electron-positron
background which withdraw transverse momentum from photons and is
polarization-dependent. If the photon frequency red-shift implies a
change in time, a clock would go faster for increasing magnetic
field intensity.
\end{abstract}

\keywords{Redshift; Photons.}

\maketitle

 \section{Introduction}
From Newton gravity theory, the action of, say, the Sun on a planet
is understood as a continuous taking of some amount of momentum,
along the line joining the planet to the Sun, keeping the planet on
a closed or open orbit, according to its initial energy. The bending
of light rays obey to a similar effect, although General Relativity
predicts twice the non relativistic shift.

In classical electrodynamics, a charged spinless particle entering
in a region where a magnetic field $B$ exists, does not vary its
energy. If we assume a particle with initial transverse momentum
$p_t$, after entering in that region, Lorentz force makes it to
rotate in such a way that its tangential momentum is equal to $p_t$.
The magnetic field withdraw momentum, (which is not conserved)) but
not energy. In the quantum case, the situation is quite different:
the energy eigenvalues of a charged particle are dependent on $B$.

A photon propagating in vacuum  in a strong magnetic field has
magnetic properties due to the dependence of $\omega$ on $B$ through
its interaction with the electron-positron virtual quanta, which is
found by solving the photon dispersion
 equation giving the photon energy dependence on the
self-energy tensor. This problem was studied by Shabad in 1975
\cite{shabad2}, who solved the dispersion equations in the limit of
frequencies close to the first pair creation threshold, where the
self-energy diverges. Paper \cite{shabad2} is a fundamental one, and
from its results one may conclude that there is a close analogy
between the bending of the light path in general relativity, a
classical problem, and the similar bending of the photon path in
magnetized vacuum, in the framework of quantum electrodynamics in
external fields. Much more, as shown in \cite{EliPRD}, the
dependence of the photon frequency on the magnetic field lead to a
(polarization dependent) paramagnetic behavior in the whole interval
of frequencies from zero to the first pair creation threshold, which
is named the region of transparency.

But photon paramagnetism implies that $\partial \omega/\partial B
<0$, and it leads to \textit{photon red-shift} in a magnetic field.
It is manifested due to the withdraw of photon transverse momentum
by the magnetized virtual electron-positron pairs. This suggests a
change in the course of time also.

\section{Electron-Positron quantum mechanics in magnetic
field}
 Let us recall some details about the
electron-positron quantum mechanics in a magnetic field. We assume
some magnetic field defined by the field invariants ${\cal
F}=2B^2>0$, ${\cal G}=0$. In a given coordinate system, a constant
uniform magnetic field $\textbf{B}$, (taken for instance, along the
$x_3$ axis), produces a symmetry breaking of the space symmetry. For
electrons and positrons ($e^{\pm}$) physical quantities are
invariant only under rotations around $x_3$ or displacements along
it \cite{Johnson}. This means that the conserved quantities, i.e.,
those commuting with the Hamiltonian operator, are all parallel to
$\textbf{B}$, as angular momentum and spin components
$J_3$,$L_3$,$s_3$ and the linear momentum $p_3$. By using units
$\hbar=c=1$, the energy eigenvalues for $e^{\pm}$ are
$E_{n,p_3}=\sqrt{p_3^2+m^2+ eB(2n+1+s_3)}$ where $s_3=\pm 1$ are
spin eigenvalues along $x_3$ and $n=0,1,2..$ are the Landau quantum
numbers. In other words, in presence of $B$, the transverse squared
energy $E_{n,p_3}^2-p_3^2$ is is given by the eigenvalues of an
oscillator-like operator added to the spin $s_3$ operator, which
quantizes in integer multiples of $eB$. The conserved quantities are
those commuting with the Hamiltonian operator and we stress that the
electron-positron transverse momentum squared $p_1^2+p_2^2$ is not a
quantum mechanical observable and is not conserved.(The
corresponding, photon momentum component $k_{\perp}^2$ is either not
conserved). For the ground state $n=0$, $s=-1$, the integer is zero.
Quantum states degeneracy with regard spin is expressed by a term
$\alpha_n=2-\delta_{0n}$, whereas degeneracy with regard to orbit's
center coordinates leads to a factor $eB$, The quantity $1/eB$
characterizes the spread of the $e^{\pm}$ spinor wavefunctions in
the plane orthogonal to $B$. The magnetic moment operator $M$  is
defined as the quantum average of $M= -\partial H/\partial B$, where
$H$ is the Dirac Hamiltonian in the magnetic field $B$, and it is
not a constant of motion. But its time-dependent terms vanish after
quantum averaging and it leads to $\bar M=-\partial
E_{n,p_3}/\partial B$.

 Due to the explicit symmetry breaking, the four momentum operator
acting on the vacuum state does not have a vanishing four-vector
eigenvalue, $P_{\mu}|0,B> \neq 0$. The components $P_{1,2}$ does not
commute with the Hamiltonian operator $H$, and are not conserved.
The $e^{\pm}$ quantum vacuum energy density is given by
$\Omega_{EH}=-eB\sum_{n=0}^\infty \alpha_n \int dp_3 E_{n,p_3}$.
After removing divergences it gives the well-known Euler-Heisenberg
expression $\Omega_{EH}=\frac{\alpha
B^2}{8\pi^2}\int_0^{\infty}e^{-B_c x/B}\left[\frac{coth x}{x}
-\frac{1}{x^2}-\frac{1}{3}\right]\frac{d x}{x}$ which is an even
function of $B$ and $B_c$, where $B_c=m^2/e\simeq 4.4\times
10^{13}$G is the Schwinger critical field. The magnetized vacuum is
paramagnetic ${\cal M}_V=-\partial \Omega_{EH}/\partial B>0$ and is
an odd function\cite{Elizabeth} of $B$. For $B<<B_c$ it is ${\cal
M}_V=\frac{2\alpha}{45 \pi}\frac{B^{3}}{B_c^2}$, where $\alpha$ the
fine structure constant. ${\cal M}_V$ is obviously be understood as
the modulus of a vector parallel to $\textbf{B}$.

\section{The photon red-shift from  Shabad's dispersion equations}
The diagonalization of the photon self-energy tensor leads to the
equations \cite{shabad2}
\begin{equation}
 \Pi_{\mu
\nu}a^{(i)}_{\nu}=\kappa_{i}a^{(i)}_{\mu},
\end{equation}
 having
three non vanishing eigenvalues
 and three eigenvectors for $i=1,2,3$, corresponding to three photon propagation modes. One additional eigenvector is
the photon four momentum vector $k_{\nu}$ whose eigenvalue is
$\kappa_{4}=0$. \cite{shabad2}.

 The dispersion equations, obtained as the zeros of
the photon inverse Green function $D^{-1}_{\mu\nu}=0, $
 after diagonalizing the polarization operator, by using the
 variables $z_1=\omega^2 - k_{\parallel}^2$, $z_2=k_{\perp}^2$, are

\begin{equation}
k^2=\kappa_{i}(z_2,z_1,B) \hspace{1cm} i=1,2,3.
\end{equation}
 After solving the dispersion equations for $z_1$ in terms of $z_2$
we get
\begin{equation}
\omega^{(i)2}=\vert\textbf{k}\vert^2+{\mathfrak M}^{2(i)}\left(z_2,
B\right) \label{eg2}
\end{equation}
We recall that for propagation  orthogonal to $B$ the mode $i=2$ is
polarized along $B$ and the $i=3$ is polarized perpendicular to $B$.
The expansion of $\kappa_{i}$ in the limit of low frequency $\omega
\ll 2m$, low magnetic field $b=B/B_c \ll 1$ was done in
(\cite{EliPRD}). By taking the first two terms in the
$\kappa_{i}^{(0)}$
 series expansion,
\begin{equation}
{\mathfrak M}^{2(2)}= -\frac{7 \alpha z_2}{45 \pi } \left(b^2
-\frac{26 b^4}{49}\right),\label{eme2}
\end{equation}
 and
\begin{equation}
 {\mathfrak M}^{2(3)}= -\frac{4 \alpha z_2}{45 \pi}
\left(b^2 -\frac{12 b^4}{7}\right).\label{eme3}
\end{equation}
Up to fields $B\sim 0.4 B_c$ one can neglect the $b^4$ term and
write as a good approximation the dispersion equation for these
modes as,
\begin{equation}
\omega^2-k^2_{\parallel}=k_{\perp}^2 \left(1- \frac{C^{i} \alpha
b^2}{45 \pi }\right) \label{Dispeq}
\end{equation}
where $C^{i}=7,4$ for $i=2,3$. Eq. (\ref{Dispeq}) can be interpreted
as the ``out" dispersion equation, where the ``in" one is light come
equation $\omega^2=k^2_{\parallel} + k_{\perp}^2$. The effect of the
magnetic field is to decrease by a factor $f(B)^{(i)}=\left(1-
\frac{C^{i} \alpha b^2}{45 \pi }\right)$ the incoming transverse
momentum, to the effective value $k_{eff \perp}^2=k_{\perp}^2
f(B)^{(i)} $. From (\ref{eme2}), (\ref{eme3}), the frequency is red
shifted  when passing from a region of magnetic field $B$ to another
of increased field $B + \Delta B$. In the same limit it is,
\begin{equation}
\Delta \omega^{(2)}=- \frac{14 \alpha z_2 b \Delta b}{45 \pi
|\textbf{k}|}<0,
\end{equation}
and

\begin{equation}
\Delta \omega^{(3)}= -\frac{8 \alpha z_2 b \Delta b}{45 \pi
|\textbf{k}|}<0.
\end{equation}

Thus, the red shift differs for longitudinal and transverse
polarizations. The magnetic field decreases the frequency of
radiation incoming to magnetized vacuum, and this decrease depends
on its polarization.

To give an order of magnitude, for instance, for photons of
frequency $10^{20}$ Hz, and magnetic fields of order $10^{12}$ G,
$|\Delta \omega| \sim 10^{-6}\omega$.

In the high frequency limit $m^2\lesssim -z_1\leq 4m^2 ,B\lesssim
B_c$, the energy gap between successive Landau energy levels for
electrons and positrons is of order close to the electron rest
energy. The photon self-energy diverges for values of $-z_1=
k_{\perp}^{\prime 2}$.

 In the vicinity of the first threshold $n=n^{\prime}=0$ and by
considering $k_\perp\neq0$ and $k_\parallel\neq0$, according to
~\cite{shabad2}, the physical eigenwaves are described by the second
and third modes, but only the second mode has a singular behavior
near the threshold. The photon redshift can be written as
\begin{equation}
\Delta \omega^{(2)}=
 \frac{(-4m^2-z_1)(1+\frac{z_2}{2eB})\alpha
m^3e^{-\frac{z_2}{2eB}}}{\omega B_c \left((4m^2+z_1)^{3/2}+b\alpha
m^3e^{-\frac{z_2}{2eB}}\right)}<0,\label{FRR1}
\end{equation}

The expression (\ref{FRR1}) has a minimum near the
threshold\cite{EliPRD}. We observe that the red shift effect in the
magnetic field acts as in the gravitational case for
\textit{increasing fields}, but in the latter, being the
gravitational potential negative, when it increases, its absolute
value \textit{decreases}. In the magnetic field case, the increase
is towards larger \textit{positive values} of the field $B$.

 The photon frequency shift in a magnetic field suggests
a variation in the course of time. This problem requires further
research, but in a simple approximation, let us assume a strongly
magnetized region of space, for instance, a star. We assume that
there is axial symmetry and in small regions, say, concentric
equipotential shells,
 the magnetic field may be considered as
constant in each one, and an increase  means to pass from a shell to
a neighbor one. Actually, if a train of waves of a fixed
polarization and frequency $\omega$, containing $n$ complete
oscillations during the time $T_1$, is sent from one shell of field
$B+\Delta B$ to the neighbor shell having field $B$, the relation
between the frequency and the interval $T_1$ is:
\begin{equation}
T_1= 2\pi n /\omega  ~\label{7}
\end{equation}
The frequency of the same train of waves at the neighbor shell can
be measured by dividing $2\pi n$ between the duration of the train.
The obtained number is $\omega' >\omega$ and it means that the
interval $T_2$  corresponding to $n$ oscillations are
\begin{equation}
T_2 = 2\pi n / \omega'         ~\label{8}
\end{equation}

From (\ref{7}), (\ref{8})  it is deduced that:
\begin{equation}
\frac{T_1 - T_2}{T_1} =|\Delta \omega|/\omega'  \label{9}
\end{equation}

Thus  $T_1\simeq(1+|\Delta \omega|/\omega)T_2$.

 That is, the clock at the first shell measured
for the duration of the wave train an interval of time greater than
in the second. Times goes faster for increasing $B$.

\section{Acknowledgments}
The authors thank OEA-ICTP for support under Net-35. One of the
authors (H.P.R.) thanks Jose Helayel-Neto for hospitality at CBPF.

\end{document}